\documentclass[runningheads]{llncs}

\usepackage[T1]{fontenc}
\usepackage{mathtools}
\usepackage{booktabs}
\usepackage{cite}
\usepackage{breakcites}
\usepackage{graphicx}
\usepackage[breaklinks=true]{hyperref}
\usepackage{color}

\hypersetup{
    colorlinks=true,
    citecolor=blue,
    linkcolor=blue,
    filecolor=blue,      
    urlcolor=blue
    }
\usepackage{eso-pic,xcolor}

\definecolor{Ocean}{cmyk}{1,0,0.2,0.55}
\definecolor{Grey}{cmyk}{0,0,0,0.6}
\definecolor{ULGray}{cmyk}{0,0,0.03,0.1}
\usepackage{dsfont}
\usepackage{amssymb}
\usepackage{amsmath, systeme}
\usepackage{amstext}
\usepackage{mathtools}
\usepackage{array,multirow}
\usepackage{siunitx}
\usepackage{gensymb}
\usepackage{algorithm}
\usepackage{algpseudocodex}
\algrenewcommand{\algorithmiccomment}[1]{\hskip3em$//$ #1}

\usepackage{subfigure}
\usepackage{academicons}

\usepackage{tikz}
\usetikzlibrary{plotmarks}

\definecolor{mygreen}{HTML}{A0BB2F}
\definecolor{myorange}{HTML}{FF8C21}
\definecolor{myyellow}{HTML}{EFCA46}
\begin{document}

\title{IM-MoCo: Self-supervised MRI Motion Correction using Motion-Guided Implicit Neural Representations}
\titlerunning{IM-MoCo}
\author{Ziad Al-Haj Hemidi \and
	Christian Weihsbach \and
	Mattias P. Heinrich}
\authorrunning{Al-Haj Hemidi et al.}
\institute{Institute of Medical Informatics, Universität zu Lübeck, Lübeck, Germany\\
	\email{z.alhajhemidi@uni-luebeck.de} \\
	\url{https://github.com/MDL-UzL/MICCAI24_IMMoCo.git}}

\maketitle

\begin{abstract}
	Motion artifacts in Magnetic Resonance Imaging (MRI) arise due to relatively long acquisition times and can compromise the clinical utility of acquired images. Traditional motion correction methods often fail to address severe motion, leading to distorted and unreliable results. Deep Learning (DL) alleviated such pitfalls through generalization with the cost of vanishing structures and hallucinations, making it challenging to apply in the medical field where hallucinated structures can tremendously impact the diagnostic outcome. In this work, we present an instance-wise motion correction pipeline that leverages motion-guided Implicit Neural Representations (INRs) to mitigate the impact of motion artifacts while retaining anatomical structure. Our method is evaluated using the NYU \emph{fastMRI} dataset with different degrees of simulated motion severity. For the correction alone, we can improve over state-of-the-art image reconstruction methods by $+5\%$ SSIM, $+5\:db$ PSNR, and $+14\%$ HaarPSI. Clinical relevance is demonstrated by a subsequent experiment, where our method improves classification outcomes by at least $+1.5$ accuracy percentage points compared to motion-corrupted images.
\end{abstract}
\section{Introduction}
\label{sec:introduction}

Magnetic Resonance Imaging (MRI) is a non-invasive imaging technique providing detailed images of soft-tissue structures. However, common artifacts, particularly motion-related ones, degrade image quality, impacting clinical diagnosis and leading to high clinical costs~\cite{slipsager2020quantifying}.

Various retrospective strategies for motion correction (MoCo) have been developed~\cite{chen2023deep, Zaitsev_Maclaren_Herbst_2015}. Retrospective correction handles motion artifacts post-acquisition, offering flexibility without requiring scanner modifications. This approach benefits from prior knowledge such as from navigators~\cite{gallichan2016retrospective}, and from the imaged object itself in so-called autofocus~\cite{atkinson1999automatic} or in data-consistency (DC)~\cite{haskell2018targeted} methods.
Unfortunately, these methods face challenges in solving poorly conditioned, non-convex optimization problems with high computational complexity. Image-based deep learning (DL) models, like convolutional neural networks (CNNs)~\cite{al2022stacked,kustner2019retrospective} and generative adversarial networks~\cite{armanious2020medgan}, have been employed for direct motion-refined image estimation to overcome the challenges mentioned earlier. While promising, image-based DL-based methods can be unstable, struggle with pathology preservation, and introduce undesired alterations or hallucinations~\cite{chen2023deep}, raising concerns about their suitability for clinical applications. Specific approaches break down the problem into intermediate steps to address instability, maintain DC, and alleviate network hallucinations. For instance, CNNs are employed to identify subsets of k-space lines affected by motion, which can either be excluded~\cite{oksuz2019detection} or down-weighted~\cite{eichhorn2023physics} in the context of a DC-based reconstruction. However, these methods do not explicitly model motion (transformations) during reconstruction, potentially compromising quality with many motion-corrupted lines. Efforts to improve efficiency integrate image-based MoCo CNNs into DC-based~\cite{haskell2019network} or autofocus~\cite{kuzmina2022autofocusing+} iterative algorithms. Alternatively, they replace the optimization-based motion estimation with a DL network and incorporate this information into the reconstruction process~\cite{Hossbach_Splitthoff_Cauley_Clifford_Polak_Lo_Meyer_Maier_2023}. However, they rely on supervised training and require large paired datasets, which are difficult to obtain in the medical domain.
Recently, Implicit neural representations (INRs) were used to model images as a function, spurring significant interest in instance-optimization-based reconstruction~\cite{10350527}. INRs have proven successful in self-supervised MRI reconstruction, addressing issues present in prior methods~\cite{feng2023imjense, huang2023neural}. Despite this, the potential of INRs for MoCo remains untapped.

In this work, we introduce a novel MoCo pipeline utilizing INRs to address motion artifacts effectively while preserving anatomical structures in MRI scans. Our contributions are 1) Developing a motion-guided INR-based method called IM-MoCo, combining a k-space line motion detection network ($k$lD-Net), an Image INR, and a motion-INR.
2) Demonstrating its efficacy on the NYU \emph{fastMRI} dataset~\cite{knoll2020fastmri,zbontar2018fastmri} through significant reductions in simulated motion-induced distortions compared to competing methods.
3) Highlighting its potential by improving pathology classification of \emph{fastMRI+} annotations~\cite{zhao2022fastmri+}. These findings position IM-MoCo as a promising advancement for enhancing MRI quality for scans with light and heavy motion corruption.

\begin{figure}[h!]
	\includegraphics[width=\textwidth]{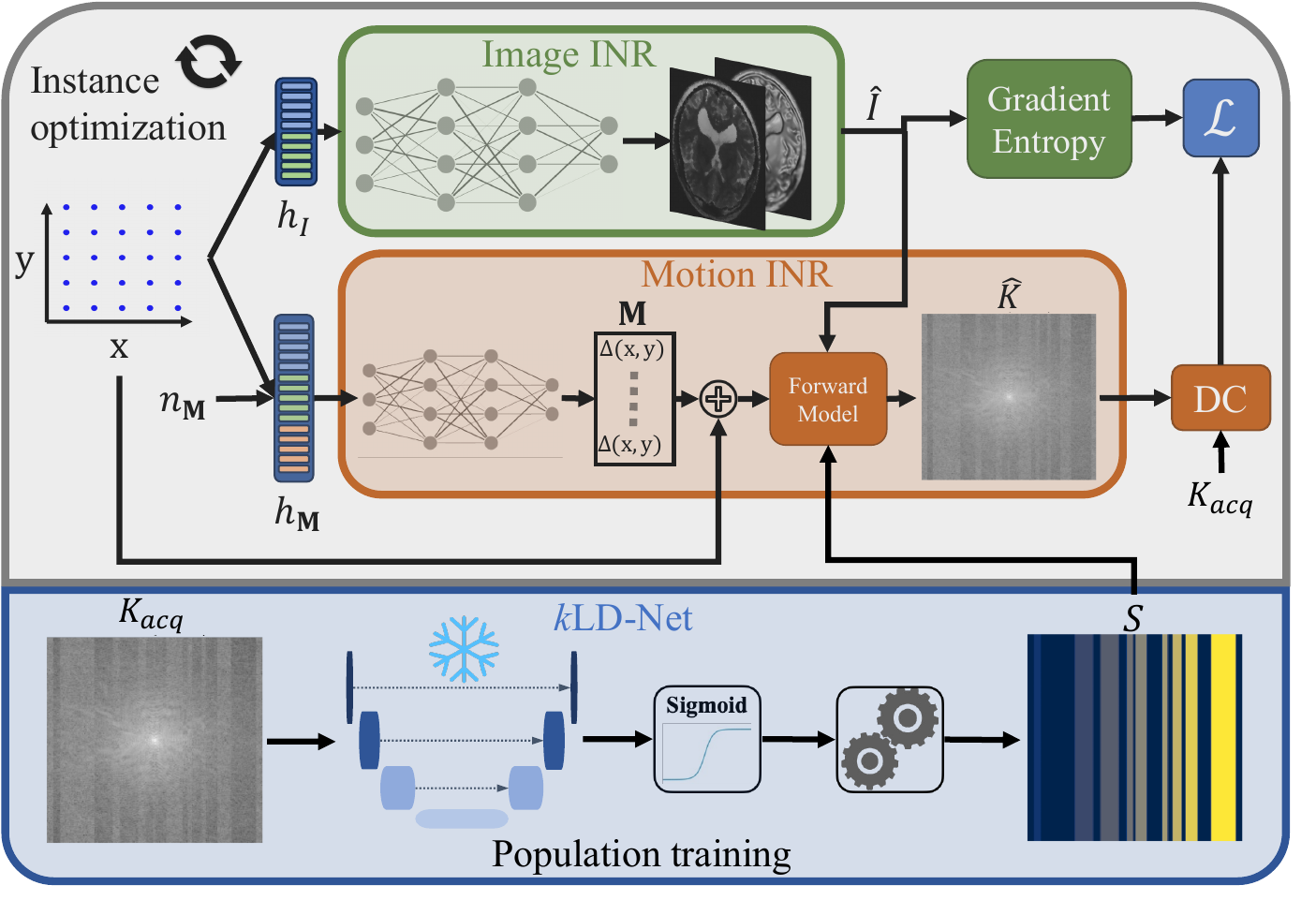}
	\caption{IM-MoCo pipeline. The pre-trained $k$lD-Net takes a motion-corrupted $k$-space and outputs a motion mask, which is post-processed to yield a list of movement groups indicated by the different colors in $S$ for better visualization. The number of movements and corresponding lines guide the Motion INR and the Image INR. The Image INR predicts the motion-free image and the Motion INR is used for guidance by optimizing the forward motion yielding a motion-corrupted $k$-space. The discrepancy between the motion-corrupted and the measured $k$-space is minimized using the data-consistency (DC) loss. The gradient entropy is a denoiser on the Image-INR output to impose crisp image priors. The final motion-corrected image is the output of the Image INR.}
	\label{fig:IM-MoCo}
\end{figure}

\section{Materials \& Methods}
\label{sec:materials_methods}
An overview of the proposed IM-MoCo pipeline is depicted in Fig.~\ref{fig:IM-MoCo}. Our pipeline comprises three steps: 1) Motion Detection with $k$LD-Net, 2) Generation of motion-free intensity priors with an Image INR, and 3) Forward application of predicted motion from a Motion INR.

\subsection{Physical Model \& Motion Simulation}
\label{sec:motionsim}

The physical forward model, $K = \mathbf{F}I + \epsilon$, of a 2D rigid motion-induced MRI acquisition $K \in \mathbb{C}^{N_x\times N_y \times N_c}$ with $N_x$ and $N_y$ being the number of points in the frequency and phase encoding directions, respectively, and $N_c$ being the number of coils, can be described as follows:
\begin{equation}
	K = \displaystyle\sum_{t=1}^T S_t \odot \mathcal{F} \mathbf{C} \mathbf{M}_t I,
	\label{eq:forwardmodel}
\end{equation}
where $I \in \mathbb{C}^{N_x\times N_y \times N_c}$ is the desired motion free image, $\mathbf{M}_t$ is the motion transform at time point $t$, $\mathbf{C} \in \mathbb{C}^{N_x\times N_y \times N_c}$ are the coil sensitivity maps, $\mathcal{F}$ is the Fourier transform, $\odot$ a sampling operator, and $S_t$ is the sampling mask at time point $t$. The motion transform $\mathbf{M}_t$ combines rotation and planar translations in $x$ and $y$ directions. The sampling mask $S_t$ is a binary mask indicating the $k$-space lines acquired at time $t$. Equation \eqref{eq:forwardmodel} describes motion in the image domain. However, the forward model can also be formulated in $k$-space as it is mathematically equivalent~\cite{loktyushin2013blind}. Irrespective of the space in which motion is described, for the simulation of motion, it is crucial to synchronize the sampling of motion $S_t$ with the MR $k$-space filling scheme to simulate realistic artifacts accurately. This work follows previously developed methods~\cite{Lee2020DeepLI} and relies on the simulation of rigid-body motion in 2D single-coil cartesian sampled $k$-spaces filled sequentially left-to-right.

\subsection{IM-MoCo}
\subsubsection{Motion Detection.}
\label{sec:motiondetection}
In the first step of our pipeline, we present a $k$-space Line Detection Network~\cite{eichhorn2023physics,oksuz2019detection} ($k$LD-Net) to detect motion-corrupted lines in raw $k$-space data following a left-to-right sequential cartesian sampling scheme. The $k$LD-Net, built on a U-Net~\cite{ronneberger2015u}, takes a complex-valued motion-corrupted $k$-space as input, where the real and imaginary parts are concatenated in the channel dimension, and outputs a binary mask predicting the motion-corrupted lines denoted as $S$ in Eq.~\eqref{eq:forwardmodel}. The four layers of the U-Net consist of convolutions with a kernel size of $3 \times 3$, batch normalization, ReLU activation, and an average pooling layer with kernel size $2\times2$. The number of channels (starting from $16$) is doubled after each down-sampling operation and halved after each up-sampling operation. A final $1\times1$ convolution outputs the predicted mask $\hat{S}$. The network is trained with the binary cross entropy loss with logits between the predicted and ground truth masks.
During inference, the raw prediction of the network is activated with $sigmoid$ and thresholded ($> 0.5$) to yield a binary prediction mask. We consider a line corrupted if $20\%$ of its frequencies are classified as so, and we generate an index list, e.g., $[1, 1, 0, 0, 0, 1, 1, 1]$, where $1$ indicates a motion-corrupted line and $0$ a motion-free one. We then post-process it to yield a list of movement groups, where each group is a list of line indices for one movement, e.g., $[[1, 1, 0, 0, 0, 0, 0, 0], [0, 0, 0, 0, 0, 1, 1, 1]]$ for the previously mentioned example. The following steps use the number of movements and corresponding lines to guide the motion correction process.

\noindent\textbf{Motion Correction.}
We use $\hat{S}$ from the $k$LD-Net (step 1) in a MoCo model, which is a combination of an Image INR (step 2) that acts as an implicit image prior and a Motion INR (step 3) to search for motion that satisfies the forward model (see Sec.~\ref{sec:motionsim}). We can rewrite the forward model $\mathbf{F}$ from Eq.~\eqref{eq:forwardmodel}, for single-coil scans, as $\mathbf{F}_{\theta, \Psi}$ as follows:
\begin{equation}
	\hat{K} = \displaystyle\sum_{t=1}^T \hat{S}_t \odot \mathcal{F} \;\text{INR}_\theta(h_{\mathbf{M}}(n_\mathbf{M},\mathbf{x}))_t \; \text{INR}_\Psi (h_I(\mathbf{x})),
	\label{eq:forwardmodel_2}
\end{equation}
where $\text{INR}_\Psi$ is the image INR with optimizable parameters $\Psi$, $h_I(\mathbf{x})$ is an encoding of the 2D coordinate $\mathbf{x}$, $\text{INR}_\theta$ is the motion INR with optimizable parameters $\theta$, $h_{\mathbf{M}}(n_\mathbf{M},\mathbf{x})$ is an encoding of the 2D coordinate $\mathbf{x}$ and $n_\mathbf{M}$, which is a sequence of linearly spaced numbers from $-1$ to $1$ representing the number of movement groups. $\hat{S}_t$ is the predicted motion mask at time $t$.
Each INR is built from MPLs taking encodings as input. To encode our data, we use hash grid encoding~\cite{muller2022instant} for $h_Is(\mathbf{x})$ and $h_{\mathbf{M}}(n_\mathbf{M},\mathbf{x})$. This encoding introduces learnable feature grids that enable a speed-up by replacing large MLPs with a look-up table of feature vectors and a much smaller MLP. The Image INR consists of three layers with $256$ channels and ReLU activations, while the Motion INR has three layers with $64$ channels and tanh activations. The $\text{INR}_\Psi$ predicts a 2-channel complex-valued intensity image, while $\text{INR}_\theta$ predicts $n$ transformation grids for $n$ movements.

In each iteration of the pipeline, we optimize the $\text{INR}_\theta$ and $\text{INR}_\Psi$ in an end-to-end fashion by feeding the encoding and the mask $\hat{S}$ through the forward model $\mathbf{F}_{\theta, \Psi}$ which in term predicts a motion-free intensity image $\hat{I}$, applies the motion to the intensity image resulting in a motion corrupted image that is transferred to the $k$-space domain by the Fourier transform yielding $\hat{K}$. Using the DC-loss in k-space, we measure the discrepancy to the acquired $k$-space $K_{acq}$. We use the gradient entropy on $\hat{I}$ for regularization in image-space. The motion correction loss is then defined as follows:
\begin{equation}
	\mathcal{L} = \underbrace{\frac{1}{N} \sum_{i=1}^{N} || K_{acq,i} - \hat{K}_i ||_2^2}_{\text{DC-loss in} \; k\text{-space}}+ \underbrace{\lambda(- \sum_{i=1}^{N}\nabla \hat{I}_i  \cdot  \log(\nabla \hat{I}_i))}_{\text{Regularization in image-space}} ,
	\label{eq:loss}
\end{equation}
where $N$ is the number of data points, $\nabla$ is the gradient operator, and $\lambda$ is the regularization weight. The regularization is used to introduce image priors to the Image INR as a denoiser, forcing the INR to learn an artifact-free image. To ensure better motion estimates, high-frequencies are suppressed at the beginning of the optimization by setting a large weight $\lambda$ and halving it every $s$ step after $i$ iterations~\cite{lin2021barf}. The model is optimized until convergence, and the final motion-corrected image is the output of $\text{INR}_\Psi$.

\section{Experimental Results}
\label{sec:experimental_results}
\subsection{Datasets \& Motion Simulation}
\label{sec:datasets}
To evaluate our method, we use the MRI $T_2$-weighted brain $k$-space data from the open-source NYU \emph{fastMRI} database~\cite{knoll2020fastmri,zbontar2018fastmri} in our first experiment. We employed $300$ 2D slices from different patients and split them into $200$/$50$/$50$ for training/validation/testing. We cropped the images to $320\times320$ and combined the coils. The second dataset contains $T_1$-weighted and FLAIR brain \emph{fastMRI} scans, which are part of the \emph{fastMRI+}~\cite{zhao2022fastmri+} annotations and were used for the downstream classification task. We classify two pathologies, \emph{Nonspecific White Matter Lesion} and \emph{Craniotomy}, resulting in $1116$ slices of $60$ subjects with a total of $2851$ annotations. We split the slices into $889$ ($1460$ annotations) for training, $174$ ($467$ annotations) for validation, and $50$ ($141$) for testing.
For all experiments, two motion scenarios were simulated~\cite{Lee2020DeepLI} with random rotations and translations between $-10$ and $+10$ in $\degree$ and $mm$, respectively. In the Light Motion scenario between $6$ and $10$ movements were introduced during simulation, whereas the Heavy Motion scenario contained between $16$ and $20$ movements.
\subsection{Experimental Setup}
\textbf{Experiment I: Motion Correction in Simulated Data.}
Here the image correction quality is assessed. We implemented the following methods: $1)$ AF~\cite{atkinson1999automatic},  $2)$ U-Net~\cite{al2022stacked}, $3)$ AF+~\cite{kuzmina2022autofocusing+}, $4)$ IM-MoCo (ours). AF is an instance-wise autofocusing approach and optimizes motion parameters using Adam for $100$ iterations. The U-Net is a population-trained model, mapping motion-corrupted images to motion-free ones, for $200$ epochs using Adam with $3e^{-4}$ learning rate and an L1-SSIM loss. AF+ combines AF and the U-Net for deep image priors, trained with default settings using the available codebase~\cite{kuzmina2022autofocusing+}. IM-MoCo is optimized for $200$ iterations using Adam with learning rates of $1e^{-2}$ for the INRs and an initial $\lambda = 1e^{-2}$, halved every $10th$ step after $100$ iterations. All methods were evaluated on the $T_2$-weighted $50$ corrupted test scans for the two motion scenarios. Results were evaluated using the structural similarity index (SSIM), peak-signal-to-noise-ratio (PSNR) and the Haar Perceptual Similarity Index (HaarPSI)~\cite{Kastryulin_Zakirov_Pezzotti_Dylov_2023}. \\
\textbf{Experiment II: Downstream Classification Task.}
%\label{sec:downstreamtask}
In this experiment, we demonstrate clinical relevance
with a classification of pathologies in motion and motion-corrected images. We extract patches of size $124\times124$ around bounding boxes of the \emph{fastMRI+} dataset and normalized between $[0,1]$. We used a pre-trained \emph{ResNet18} (PyTorch Hub\footnote[2]{\url{https://pytorch.org/hub/pytorch_vision_resnet/}}) backbone for feature extraction and added two trainable linear layers as the classification head. Using the SGD optimizer, we trained the model for $100$ epochs with a batch size of $20$ and a learning rate of $1e^{-3}$. The trained model is then applied to the test set. We report the SSIM, PSNR, and HaarPSI for extracted patches around the bounding boxes and accuracy for the classification.\\
\textbf{Implementation Details.} The $k$LD-Net was population-trained with Adam (learning rate of $1e^{-4}$) for $4200$ epochs and a batch size of $4$. We used PyTorch for implementations and an NVIDIA GeForce RTX 2080 TI with 12 GB of VRAM. The \href{https://github.com/MDL-UzL/MICCAI24_IMMoCo.git}{code} is publicly available.

\subsection{Results}
\label{sec:results}
\begin{figure}[h!]
	\centering
	\includegraphics[width=\textwidth]{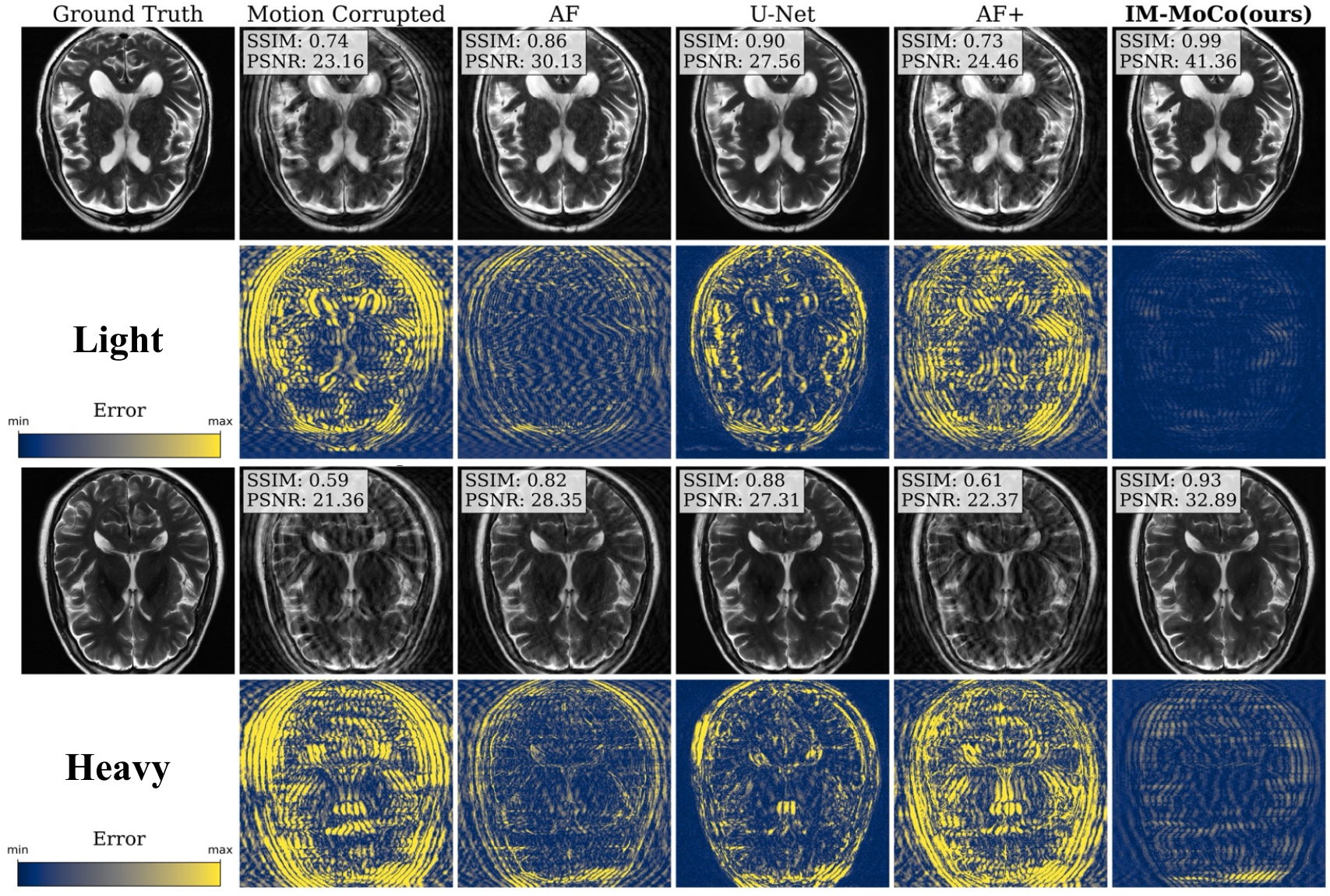}
	\caption{The visualization shows the median results of motion-corrected images of our IM-MoCo pipeline besides motion-corrupted, ground truth, and comparison methods. The first and third rows show the light and heavy correction results, respectively. The second and fourth rows show the residual error images.}
	\label{fig:motioncorrectionresults}
\end{figure}

\textbf{Experiment I: Motion Correction in Simulated Data.}
Table~\ref{tab:correction} shows quantitative results from the motion correction experiment. For light and heavy motion, corrupted images yield SSIM values of $87\%$ and $74\%$, PSNR values of $28\:db$ and $24\:db$, and HaarPSI values of $70\%$ and $56\%$, respectively. AF and U-Net improve SSIM by $7\%$ and $10\%$, PSNR by $5\:db$ and $3\:db$, and HaarPSI by $18\%$ and $16\%$, respectively. However, AF+ worsens results by ca. $2\%$ across all metrics. Compared to the best baseline, AF, our method improves SSIM/HaarPSI by $>4\%$ for SSIM, $>9\%$ for HaarPSI, and $>10\:db$ for PSNR. A qualitative comparison is shown in Fig.~\ref{fig:motioncorrectionresults}.

\begin{table}[h!]
	\footnotesize
	\vspace{1ex}
	\centering
	\caption{\label{tab:correction} Quantitative results of experiment I: motion correction for all methods and motion scenarios. We report the results as mean $\pm$ standard deviation over the test set. The arrows indicate the direction of improvement.}
	\resizebox{\textwidth}{!}{
		\begin{tabular}{@{\extracolsep{4pt}}llccc}
			\toprule
			\textbf{Scenario} & \textbf{Method}
			                  & \textbf{SSIM$\uparrow$ } & \textbf{PSNR (db)}$\uparrow$ & \textbf{HaarPSI$\uparrow$}                              \\ \midrule
			\textbf{Light}    &
			%\midrule        
			Motion corrupted  & $87.26 \pm 4.42 $        & $28.34 \pm 2.97$             & $70.48 \pm 8.69$                                        \\
			                  & AF~\cite{atkinson1999automatic}                       & $94.47 \pm 2.06$             & $33.91 \pm 2.37$           & $ 88.49 \pm 4.11$          \\
			                  & U-Net~\cite{al2022stacked}                    & $91.39 \pm 2.14$             & $30.58 \pm 2.33$           & $81.58 \pm 4.49$           \\
			                  & AF+~\cite{kuzmina2022autofocusing+}                     & $85.18 \pm 4.75$             & $27.93 \pm 2.79$           & $70.82 \pm 8.83$           \\
			                  & IM-MoCo (ours)           & $\mathbf{98.25 \pm 1.25}$    & $\mathbf{40.06 \pm 3.33}$  & $\mathbf{97.20 \pm 4.05}$  \\
			\midrule
			\textbf{Heavy}    &
			%\midrule
			Motion corrupted  & $74.06 \pm 6.36$         & $24.28 \pm 2.50$             & $56.56 \pm 8.68$                                        \\
			                  & AF~\cite{atkinson1999automatic}                       & $87.19 \pm 3.51$             & $29.84 \pm 2.32$           & $78.91 \pm 6.21$           \\
			                  & U-Net~\cite{al2022stacked}                    & $84.55 \pm 3.63$             & $27.40 \pm 2.34$           & $72.72 \pm 5.69$           \\
			                  & AF+~\cite{kuzmina2022autofocusing+}                      & $70.61 \pm 8.30$             & $24.12 \pm 2.67$           & $56.91 \pm 10.35$          \\
			                  & IM-MoCo (ours)           & $\mathbf{92.77 \pm 3.59}$    & $\mathbf{33.06 \pm 3.59}$  & $\mathbf{ 87.29 \pm 9.38}$ \\
			\bottomrule
		\end{tabular}
	}
\end{table}

\noindent\textbf{Experiment II: Downstream Classification Task.}

The results are summarized in Table~\ref{tab:dtection}. Motion-free images achieve an accuracy of $97\%$. For corrupted images, the accuracy is $96\%$ for light and $94\%$ for heavy motion, which is due to image quality loss as reflected in SSIMs of $89\%$ and $77\%$, respectively. The U-Net's accuracy is $90\%$ for light and $88\%$ for heavy while its SSIMs are $87\%$ and $79\%$. IM-MoCo achieves an accuracy of $97\%$ for light and $96\%$ for heavy motion, with the highest SSIMs of $99\%$ and $95\%$, respectively.

\begin{table}[h!]
	\footnotesize
	\vspace{1ex}
	\centering
	\caption{\label{tab:dtection} Quantitative results of experiment II: image quality and classification accuracy improvements in patches of the \emph{fastmri+} annotations. The arrows indicate the direction of improvement.}
	\resizebox{\textwidth}{!}{
		\begin{tabular}{@{\extracolsep{4pt}}llcccc}
			\toprule
			\textbf{Scenario} & \textbf{Method}
			                  & \textbf{SSIM$\uparrow$ } & \textbf{PSNR (db)}$\uparrow$ & \textbf{HaarPSI$\uparrow$} & \textbf{Accuracy$\uparrow$}                    \\ \midrule
			Motion-free       & n.A.                     & n.A.                         & n.A.                       & n.A.                        & $97.16$          \\
			\midrule
			\textbf{Light}    &
			%\midrule        
			Motion corrupted  & $89.93 \pm 4.67$         & $28.29 \pm 4.07$             & $76.12 \pm 9.95$           & $96.32$                                        \\
			                  & U-Net~\cite{al2022stacked}                     & $87.37 \pm 4.31$             & $25.80 \pm 2.61$           & $70.87 \pm 7.74$            & $90.44$          \\
			                  & IM-MoCo (ours)           & $\mathbf{99.00 \pm 1.82}$    & $\mathbf{44.82 \pm 6.44}$  & $\mathbf{97.33 \pm 5.56}$   & $\mathbf{97.06}$ \\
			\midrule
			\textbf{Heavy}    &
			%\midrule
			Motion corrupted  & $77.03 \pm 5.74$         & $23.56 \pm 2.18$             & $58.87 \pm 6.20$           & $94.12$                                        \\
			                  & U-Net~\cite{al2022stacked}                     & $79.45 \pm 4.50$             & $23.70 \pm 2.18$           & $59.82 \pm 5.88$            & $88.24$          \\
			                  & IM-MoCo (ours)           & $\mathbf{95.26 \pm 3.31}$    & $\mathbf{34.56 \pm 5.61}$  & $\mathbf{88.48 \pm 7.80}$   & $\mathbf{96.32}$ \\
			\bottomrule
		\end{tabular}
	}
\end{table}
\section{Discussion \& Outlook}
\label{sec:discussion}
Experiment I demonstrated that our method effectively enhanced the quality of motion-corrupted images, surpassing comparison methods, which is mainly contributed by the implicit image priors and DC with the acquired $k$-space in the forward motion model. Notably, our approach exhibited robustness in addressing heavier motion scenarios as evidenced in Fig.~\ref{fig:motioncorrectionresults} and Table~\ref{tab:correction}. However, the efficacy of our method hinges on the accurate detection of the $k$lD-Net, which is contingent upon the acquisition sequence and motion simulation pattern. In Experiment II, we showcased the performance enhancement of a classification task compared to both corrupted and U-Net-corrected images as indicated in Table~\ref{tab:dtection}. This improvement in classification accuracy can be attributed to the prevention of overfitting on "healthy" features in instance-optimization methods in contrast to population training ones like the U-Net.
Although our work primarily revolved around 2D rigid motion, we are confident in the method's potential for extension to more complex motion and multi-coil data. This can be achieved by incorporating coil sensitivity map estimation following the third step of the pipeline. Future investigations will explore the utilization of different sampling patterns, such as radial sampling, and the adaptation of the motion detection network to these patterns or pattern-independent methods. While real motion cases were not tested in this study, we are optimistic regarding the adaptability of our method through adjustments to the motion detection network.

\section{Conclusion}
\label{sec:conclusion}
We have introduced IM-MoCo, a pipeline that utilizes motion-guided INRs to mitigate the impact of motion artifacts in MRI scans. We tested our method on the \emph{fastMRI} dataset and observed an improvement in image quality and pathology classification performance. Our results suggest that IM-MoCo is a promising solution for enhancing the quality of MRI scans in the presence of challenging motion artifacts. In the future, we plan to investigate the applicability of our method to real motion to validate its potential for clinical applications.

\bibliographystyle{splncs04}
\bibliography{mybibliography}

\newpage
\appendix
\section{Supplementary Material}
\label{label:supplement}

\begin{figure}[h!]
	\center
	\includegraphics[width=0.6\textwidth]{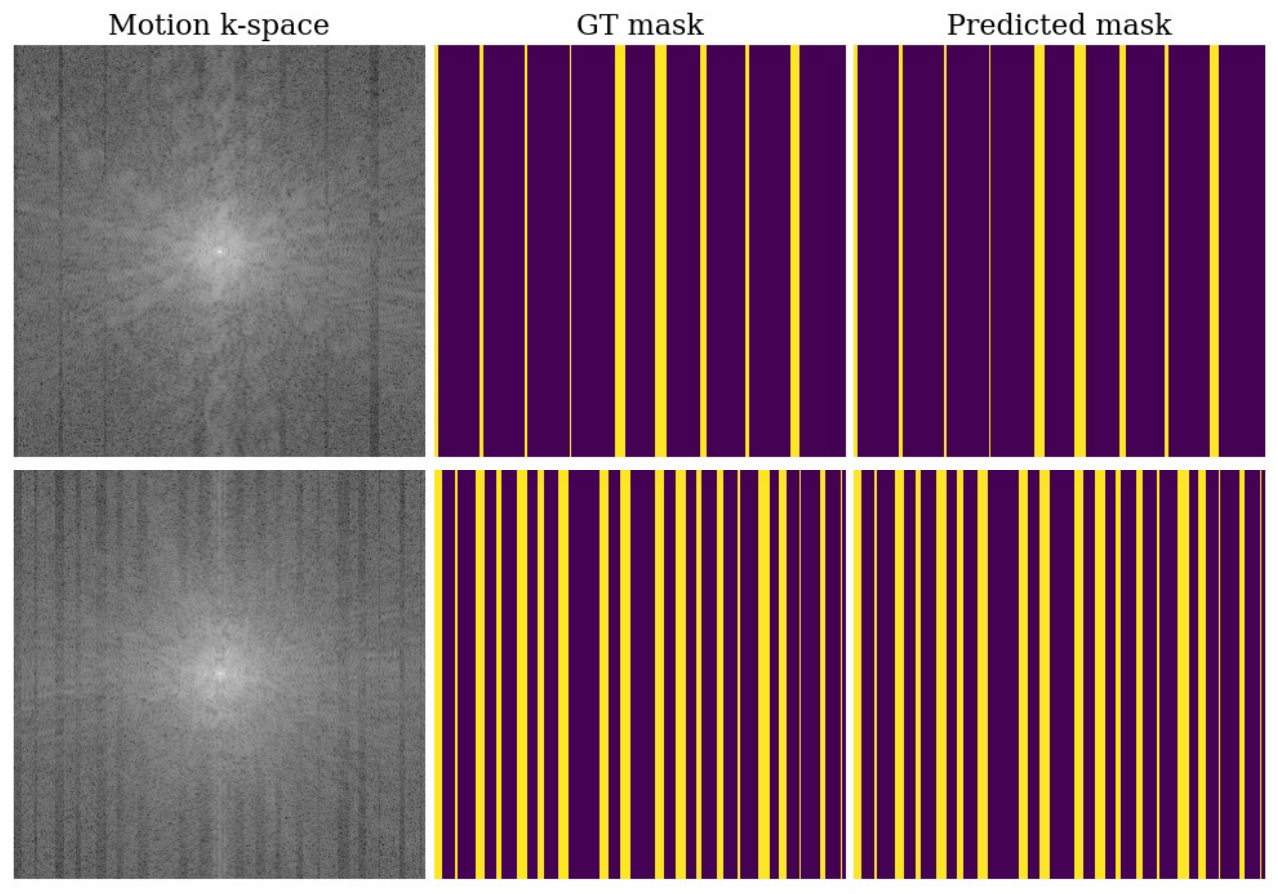}
	\caption{$k$lD-Net detection results. The first and second show the detection performance of the $k$LD-Net for light and heavy motion, respectively.}
\end{figure}
\begin{figure}[h!]
	\center
	\includegraphics[width=.9\textwidth]{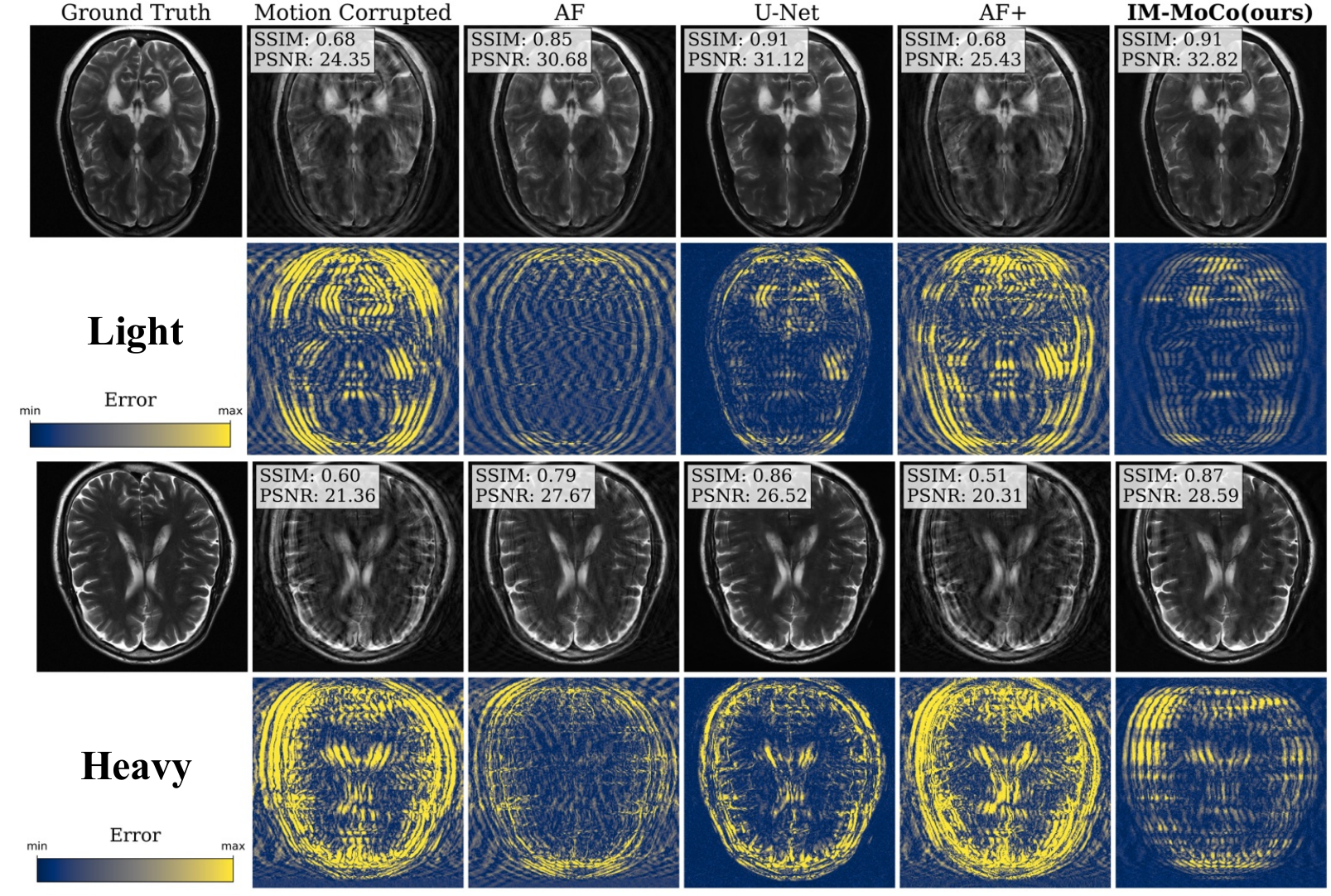}
	\caption{The visualization shows the worst results of motion-corrected images of our IM-MoCo pipeline besides motion-corrupted, ground truth, and comparison methods. The first and third rows show the light and heavy correction results, respectively. The second and fourth rows show the residual error images.}
\end{figure}

\begin{figure}[h!]
	\center
	\includegraphics[width=.9\textwidth]{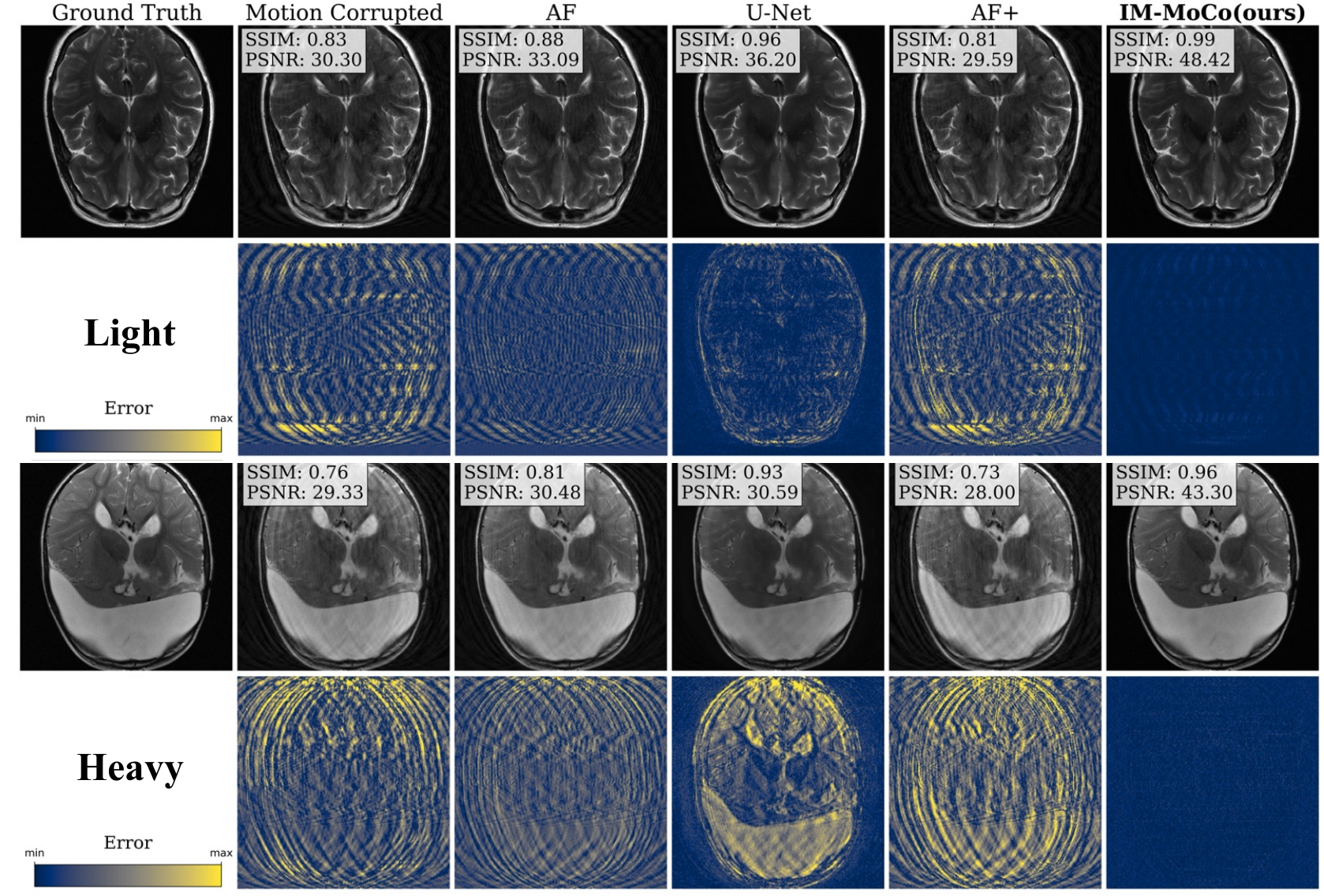}
	\caption{The visualization shows the best results of motion-corrected images of our IM-MoCo pipeline besides motion-corrupted, ground truth, and comparison methods. The first and third rows show the light and heavy correction results, respectively. The second and fourth rows show the residual error images.}
\end{figure}
\begin{figure}[h!]
	\includegraphics[width=\textwidth]{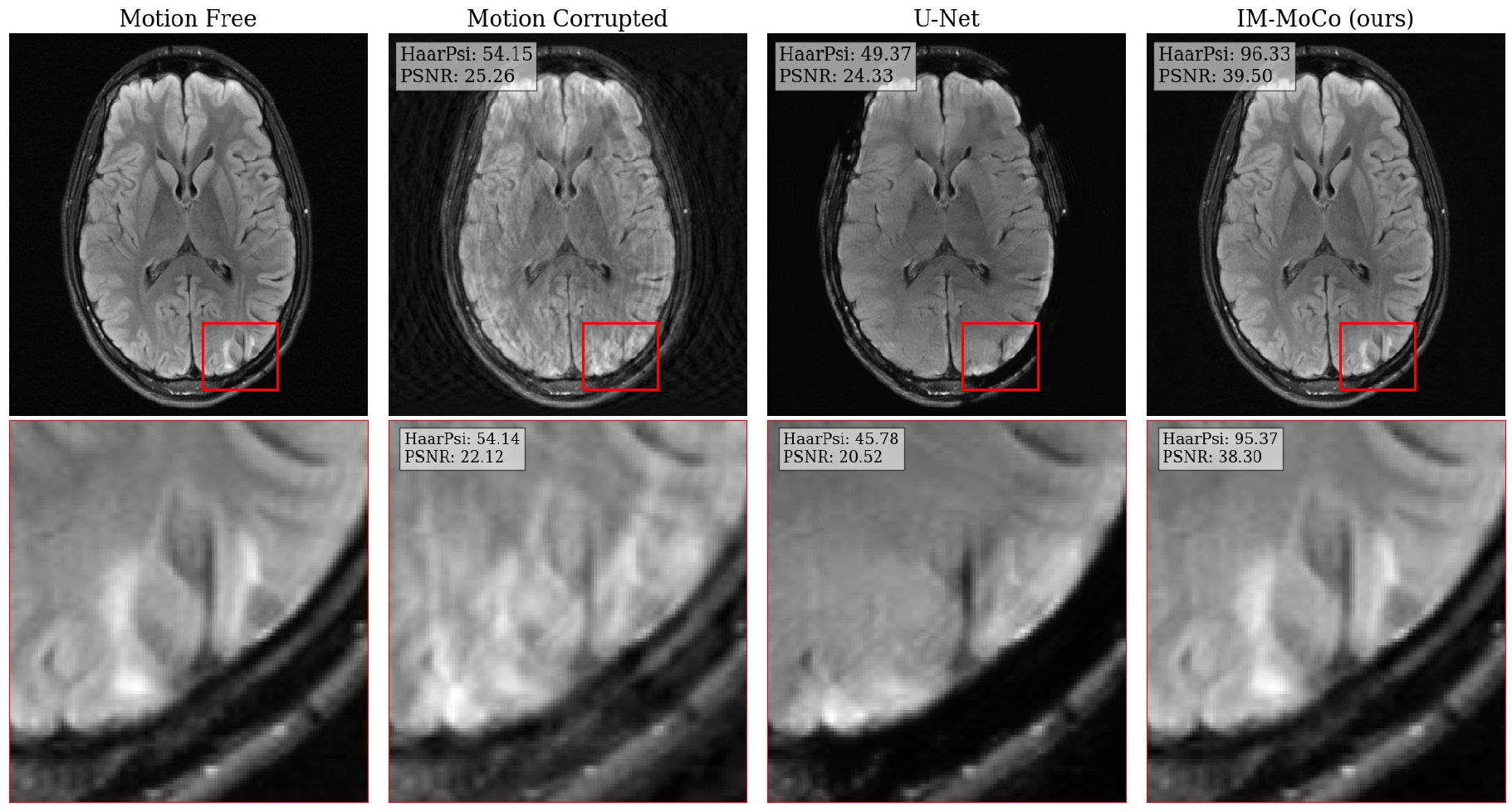}
	\caption{A visualization of a \emph{Non-specific white matter lesion} as an example. The first and second rows show the full image and the extracted patches, respectively.}
\end{figure}

\end{document}